\documentclass[aps,twocolumn,pre,floatfix]{revtex4}

\usepackage{graphicx}
\usepackage{amsmath}
\usepackage{color}

\newcommand{\figurewidth}{\columnwidth}

\newcommand{\upd}{{\rm d}}
\newcommand{\kT}{{k_{\rm B}T}}
\newcommand{\la}{\langle}
\newcommand{\ra}{\rangle}
\newcommand{\half}{{\frac{1}{2}}}
\newcommand{\fe}{f_{\rm e}}
\newcommand{\el}{{e_{\rm loop}}}
\newcommand{\eb}{{e_{\rm link}}}
\newcommand{\mup}{{\mu_{\rm p}}}

\begin{document}

\title{The Immunity of Polymer-Microemulsion Networks}
\author{Guy Hed and S. A. Safran}
\affiliation{Department of Materials and Interfaces,
Weizmann Institute of Science, Rehovot 76100, Israel}

\begin{abstract}
The concept of network immunity, {\it {i.e.}}, the robustness of
the network connectivity after a random deletion of edges or
vertices, has been investigated in biological or communication
networks. We apply this concept to a self-assembling, physical
network of microemulsion droplets connected by telechelic
polymers, where more than one polymer can connect a pair of
droplets. The gel phase of this system has higher immunity if it
is more likely to survive ({\it{i.e.}}, maintain a
macroscopic, connected component) when some of the polymers are
randomly degraded.  We consider the distribution
$p(\sigma)$ of the number of polymers between a pair of
droplets, and show that gel immunity decreases as the variance
of $p(\sigma)$ increases. Repulsive interactions between the
polymers decrease the variance, while attractive interactions
increase the variance, and may result in a bimodal $p(\sigma)$.
\end{abstract}

\maketitle

\section{Introduction}
The immunity of networks refers to  the ability of a network to preserve its
function, even when a certain number of edges or vertices is removed.
This attribute is of interest in many types of networks \cite{AlbertRMP}.
The immunity of genetic, cellular
networks  reflects the robustness of the network under a mutation or
deletion of one of the genes \cite{Wagner00}. The robustness of the metabolic
network of single cell organisms is demonstrated by their persistence and growth
despite environmental interventions \cite{Albert00}.
In communication networks, immunity refers to the ability of the network to
pass messages among most of its vertices even after some communication lines
are cut, or some vertices are damaged \cite{Cohen03}.
A communication network is immune to such an attack, if it retains
a connected section that spans enough vertices to allow some communication
(even if quite indirect) between
any two vertices.

In this paper, we extend these ideas to analyze the immunity of a
network composed of droplets connected by polymers \cite{Porte}.
The gel phase of such a system is characterized by a connected
part of the network that spans the entire volume. The immunity
of the gel is characterized by the fraction of polymers that can
be removed (by chemical breaking, depolymerization or by shear
forces) from the network while the gel maintains its macroscopic
characteristics, and does not transform into the
fluid phase. That is, a gel is immune if the physical network
that comprises the gel retains a macroscopically connected
component even though a relatively high fraction of the polymers
that form the network have been deleted.

A simple polymer gel is a network of connected
polymers in which each edge ({\it{i.e.}}, bond) of the network
consists of one polymer; in this case, the concept of immunity
is idential with that of percolation \cite{Gottlieb}. However,
in the system of polymers that join microemulsion droplets, each
edge ({\it{i.e.}}, bond) that connects two oil droplets may
contain more than one polymer -- and sometimes  many polymers.
The percolation problem deals with the removal of entire edges
from the network, while immunity refers to the removal of
individual polymers from a given edge in systems where there can
be many polymers in an edge.

Examples of equilibrium, network-forming systems include surfactant
solutions, gels of biological molecules or synthetic polymers.
A particularly elegant
experimental realization of a transient, self-assembling, physical
network has been reported in \cite{Porte}.
The system consists of oil-in-water microemulsion droplets
connected by telechelic polymers;
the latter have a hydrophilic backbone with a
hydrophobic group at each chain end.
The insolubility of these groups in the continuous water phase
and their solubility in the oil droplets
results, in some cases, in a network of droplets connected
by the polymers.
Mixtures of telechelic
polymers and microemulsions have a wide range of technological
applications, including paints, cosmetics and enhanced oil
recovery. Precise control of the structural and rheological
properties of such materials is essential for effective and reliable
performance. Such control is possible using the
telechelic additives, that form a transient network with
controlled rheological and structural properties
that depends only on the physical properties of the system such
as the size and concentration of the polymers and the droplets \cite{Porte}.

Apart from its applied interest, the telechelic-microemulsion mixtures
serve as model systems for the understanding of a
more general class of transient networks. The advantage of this
particular system is that the parameters that control the
thermodynamics and structure can be easily identified and
independently controlled: the {\em concentration of possible
vertices} (the droplets) and the {\em connectivity of the network}
(the number of polymers per droplet).

The experimental systems \cite{Porte} exhibit a phase
transition from a single phase to coexisting high density and low
density network phases. Zilman et al. \cite{Zilman03} showed that
this transition is expected from entropic considerations alone,
and occurs even without any particular energetic
interactions between the droplets and polymers. In this paper,
we extend the analysis of \cite{Zilman03} and examine the
immunity of such polymer gels to random degradation of the
polymers such as those that can result from chemical degradation
or from shear forces. As the polymers are removed, the
macroscopic network becomes disconnected and the gel reverts to
a liquid. We show the interactions between the hydrophobic
groups at the ends of the polymers have an important affect on
the immunity and the phase behavior of the system.

The relation between the degradation of the polymers and
the breakdown of the network is not trivial. Each edge in the network may
contain several polymers, each of which connect one pair of droplets.
The fraction of polymers that must be degraded in order for the network
to lose its connectivity
depends on the distribution, $p(\sigma)$, of the number of polymers in an edge.
As the variance of the distribution increases, the network becomes
more sensitive to polymer degradation, and will lose its connectivity
when a smaller fraction
of polymers is degraded.

We consider the distribution, $p(\sigma)$, for various physical situations.
Since we are treating relatively rigid and short polymers, we do not consider
the interactions between the chains, but focus on the interactions between
hydrophobic groups at the end of the polymers, which we denote as
\emph{stickers}. In particular, we predict the effect of both repulsive
and attractive interactions between the stickers within a droplet.
Repulsive interactions are
expected from entropic and excluded volume considerations,
while attractive interactions can arise
from Van der Waals forces, or from the local change of curvature that the
hydrophobic ends induce on the droplet surface.
We find that repulsive interactions decrease the variance of $p(\sigma)$
while attractive interaction increase the variance.

Attractive interactions between the stickers may
lead to the co-existence of dense edges, with high density of
polymers, with dilute edges, with low polymer density. In that
case, the distribution of the edge occupation number,
$p(\sigma)$, is bimodal, and its width is of the order of its
average. As the system approaches the critical point associated
with the emergence of this two-phase behavior, the fluctuations
in the number of  polymers in an edge as well as the variance of
$p(\sigma)$, increase. Thus, as we {\em increase} the strength of
the interaction between the stickers toward its critical value,
local concentration fluctuations increase, and the immunity {\em
decreases}.

In Sec. \ref{sec1} we describe the underlying physical model for the
network forming systems of interest, and the simplified theoretical
model we use in our calculations.
Next, in Sec. \ref{sec2}, we  examine some simple distributions for
$p(\sigma)$ and demonstrate the dependence of the immunity of the gel
on the variance of the distribution.
The effects of repulsive and attractive
interactions between the polymer ends on or in  a droplet
are presented in Sec. \ref{sec3}.
We use an expression for
the Landau free energy of the system of both
interacting polymers and droplets to obtain the effect of the
interactions on the variance
of $p(\sigma)$, and we predict the phase diagram of this system.

\section{The model} \label{sec1}

We map the physical space to a discrete regular lattice \cite{Zilman03},
where each site can be occupied by an oil droplet. Each pair of neighboring
droplets may be connected by polymers. The set of polymers that connect such
a pair is referred to as an \emph{edge} of the network, whose vertices are
the droplets.

From the applications of percolation theory to
gelation we know that if the fraction, $\kappa$, of occupied
edges exceeds the percolation threshold, $\kappa_c$, of the
lattice, the system is in the gel phase. In the usual
discussions of gel connectivity and percolation, each edge is
occupied by one polymer. After the degradation of a fraction $q$
of the polymers, the condition for the survival of the gel phase
is $\kappa(1-q)>\kappa_c$.

However, the system we discuss here has two important and interesting
differences from the usual percolation models:
\begin{enumerate}
\item The vertices of the polymer network only exist at those lattice sites
at which the oil droplets are present,
so only two neighboring vertices of the lattice
that are each occupied by droplets can be connected by an edge.
In the following, we refer to such a pair as an {\em edge} even
it is not connected by polymers, in which case we consider it as a
disconnected edge.
If a given droplet has a neighboring vacant lattice site, we refer to the
the droplet surface that face this site as a \emph{free face}.
\item An edge may be occupied by a number, $\sigma$, of polymers, whose value
is determined by volume limitations (in the case of excluded volume
interactions), explicit interactions between the polymers or between the
stickers, and the polymer chemical potential - which is fixed by their
concentration in the solution.  The polymer chemical potential determines
the average number of polymers in an edge, $\langle \sigma \rangle$.
\end{enumerate}

\section{Network immunity and edge occupation distribution} \label{sec2}

In this section, we concentrate on the effect of the edge
occupation distribution, $p(\sigma)$, on the immunity.
We do not consider here the droplet-related degrees
of freedom, but rather assume that they are fixed in a given
(static) configuration. The following section considers the
droplet-related degrees of freedom. The only degrees of freedom
we relate to in this section are occupation, $\sigma$, of the
edges between neighboring droplets.

If the droplets are dense enough, and most of the edges are
connected by one or more polymers, then the system
has a connected component that spans its entire volume,
 and the system is in the gel phase. If the fraction of
disconnected edges ({\it i.e.} those that contain no polymer),
is large, the system is in the fluid phase.   A system in the
gel phase that undergoes a degradation of a fraction of its
polymers, has higher immunity if fewer edges are completely
disconnected by the degradation process. This will depend on the
number of polymers per edge and on the variation of this number
throughout the network.

For example, if the system has
a distribution of edges in which some edges have large numbers of polymers
and some have very few, it will have lower immunity than a system in which
the number of polymers in each edge is the same.  This is because
the random removal of polymers from edges with lower than average occupancy
more readily leads to the complete disconnection of that edge,
as demonstrated in Fig. \ref{fig1}.

The probability for a given edge to be occupied by $\sigma$ polymers
is $p(\sigma)$; the fraction of unoccupied, disconnected, edges is given by
$p(0)$ - the probability of a pair of neighboring droplets to have
no polymers linking them.

After a fraction $q$ of the polymers is degraded, the new occupation of
an edge which had $\sigma$ polymers is proportional to
the product of the probability of a number, $\sigma'$ of polymers to remain
in the edge, $(1-q)^{\sigma'}$ and the probability
of the remaining polymers to be removed, $q^{(\sigma-\sigma')}$.
The resulting  distribution can be written
\begin{equation}
\tilde p(\sigma') = \sum_{\sigma=\sigma'}^\infty p(\sigma) \binom{\sigma}{\sigma'}
q^{\sigma-\sigma'} (1-q)^{\sigma'} \;.
\end{equation}

The new fraction of unoccupied edges is
$\tilde p(0) = \sum_\sigma p(\sigma) q^\sigma$. The higher this quantity for
a given initial average edge occupation, $\la\sigma\ra$, and a given
degradation probability, $q$, the lower the immunity of the system.

\subsection{Simple distributions}
In order to elucidate relation between the variance of the
edge occupation distribution $p(\sigma)$ and the immunity of the gel, we
first study two simple distributions. The first is a uniform distribution,
\begin{equation}
p(\sigma) = \left\{ \begin{array}{cc}
  \frac{1}{2s+1} & \ \ \ m-s \leq \sigma \leq m+s \\
  0 & {\rm otherwise} \end{array} \right. \;,
\end{equation}
with an average $m$ and variance $S^2_\sigma = (s^2 + s) / 3$.

If we randomly delete a fraction $q$ of the polymers, then the fraction
of unoccupied edges is
\begin{equation}
\tilde p(0) = \frac{q^{m-s}}{2s+1} \cdot \frac{1-q^{2s+1}}{1-q} \;,
\end{equation}
which is an increasing function of $s$ for any value of $q$. Thus,
in a distribution with a larger variance, the probability of an
edge to remain connected is smaller, and it is \emph{less immune}
to random polymer degradation.

\begin{figure}
\includegraphics[width=\figurewidth]{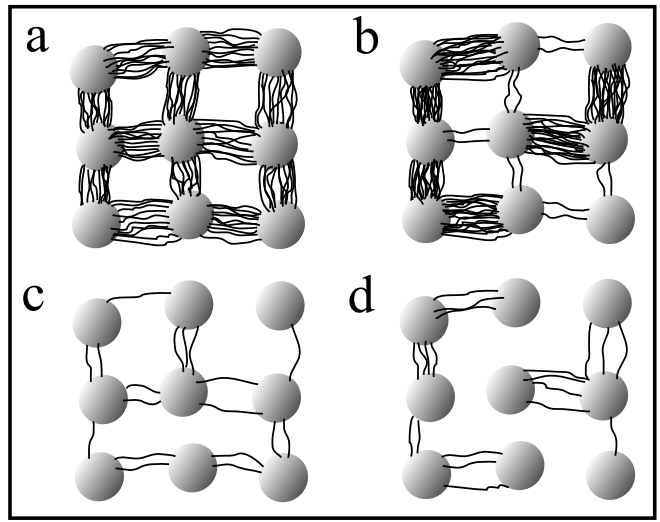}
\caption{Models of polymer-microemulsion systems
with different distributions, $p(\sigma)$, of the number of
polymers in an edge. We compare a homogeneous system (a) with 11
polymers in each edge (mean $m=11$ and distribution width $s=0$)
vs. a system with bimodal $p(\sigma)$ (b), as given by Eq.
\ref{eqtoybin} ($m=11$, $s=9$). After a random deletion of a
fraction, $q=0.85$, of the polymers, most of the edges in the
homogeneous system (c) survive, and it still retains a
macroscopically connected component. In the inhomogeneous
system, most of the dilute edges are eliminated by the deletion
(d) and the system no longer has a macroscopically connected
component.}
\label{fig1}
\end{figure}

The second distribution we would like to consider is the bimodal
distribution depicted in Fig. \ref{fig1}:
\begin{equation}\label{eqtoybin}
p(\sigma) = \left\{ \begin{array}{cc}
  1/2 & \ \ \ \sigma=m\pm s \\
  0 & {\rm otherwise} \end{array} \right. \;,
\end{equation}
with an average $m$ and variance $s^2$. After a deletion of a fraction
$q$ of the polymers, the fraction of disconnected edges is
\begin{equation}
\tilde p(0) = q^m \cosh(s\log q) \;.
\end{equation}
As the variance $s$ increases, the system become less immune to
random polymer degradation, as demonstrated in Fig. \ref{fig1}.
This is relevant to polymer-microemulsion systems with a bimodal
distribution of the number of polymers in an edge, as expected when
the attractive interactions between the hydrophobic stickers at the polymer
ends are large (see Sec. \ref{sec3}).

In the following sections we consider several physical models
corresponding to different physical interactions between the
polymers and stickers, that yield different edge occupation distributions,
$p(\sigma)$.  We calculate the variance for each situation
to estimate the gel immunity and how it depends on the interactions.

\subsection{Non-interacting polymers}
We first  examine a naive, but instructive, model where any edge can
accommodate an infinite number of polymers.
In order to compare it with other models, we
express the variance as a function of the polymer concentration, which
is the control parameter in most of the experiments \cite{Porte}.
First, we calculate the distribution in the grand canonical ensemble,
in which  the chemical potential, $\mup$, of the polymers is fixed. We then
calculate the dependence of  $\mup$ on  the polymer concentration,
to determine $\mup$.

The distribution is given by
\begin{equation}\label{eq:psignaive}
p(\sigma) =  e^{-e^{-\mup}} \frac{e^{-\sigma\mup}}{\sigma!} \;,
\end{equation}
where the factorial in the denominator comes from the fact that the polymers
are indistinguishable.
This distribution has an average and a variance
$\la\sigma\ra=S_\sigma^2=e^{-\mup}$.


\subsection{Excluded volume interactions} \label{ssev}
In a more realistic model, the excluded volume interactions
 limit the maximum
number of polymers that can reside in a given droplet.
There is thus  a finite, maximal number of polymers, $\sigma_M$,
associated with one edge.
The volume restriction may arise from Helfrich-like,
excluded volume interactions between the
polymer chains, or from the finite area
occupied by each sticker in or on the surface of a droplet.

If the only interactions between the polymers and between the stickers are
short range excluded volume interactions, we can assume that there are
$\sigma_M$ sites in an edge, that can each be occupied by one
polymer. The distribution of the edge occupation is
\begin{equation}
p(\sigma)=Z^{-1} \binom{\sigma_M}{\sigma} e^{-\sigma\mup} \;,
\end{equation}
where $Z=(1+e^{-\mup})^{\sigma_M}$ is the partition function.

The free energy per site (we take all energies in units of $\kT$)
is $\fe=-\log(Z)/\sigma_M$.
The average edge occupation is
\begin{equation}
\la\sigma\ra = \sigma_M \frac{\partial \fe}{\partial\mup}
= \frac{\sigma_M}{1+e^\mup} \;,
\end{equation}
and the variance of the distribution $p(\sigma)$ is
\begin{equation}
S_\sigma^2 = - \sigma_M \frac{\partial^2 \fe}{\partial\mup^2} =
\frac{\sigma_M}{4\cosh^2(\mup/2)} =
\la\sigma\ra \left(1-\frac{\la\sigma\ra}{\sigma_M}\right) \;.
\end{equation}
The variance here is smaller than in the case of non-interacting polymers
(Eq. \ref{eq:psignaive}). In the non-interacting
case, the variance was equal to $\la\sigma\ra$,
where in the present case it is smaller by a factor of
$1-\la\sigma\ra/\sigma_M$. As expected, the difference between the
distributions vanishes for the case $\sigma_M \gg \la\sigma\ra$.
The narrowing of the distribution by the excluded volume interactions
increases as the average number of polymers per edge approaches $\sigma_M$,
where the variance vanishes and the distribution approaches a delta function
as $\sigma \rightarrow \sigma_M$.
This  suggests that repulsive interactions between the polymers or  the
stickers increase the immunity of the gel under a random degradation of the polymers.

\section{Interacting stickers} \label{sec3}

We now calculate the free energy of
a system of droplets and polymers in which the polymers show attractive
or repulsive interactions in the multi-canonical ensemble, where the
number of polymers and droplets is not fixed, and is determined by
the chemical potentials $\mup$ and $\mu_{\rm d}$, respectively.
We shall consider the possibility of
polymers that connect neighboring droplets as well as polymers
that form loops, in which both sticker ends reside in the same droplet.

The physical model can be associated with the vertices (droplets)
and edges of  an underlying graph. The vertices are given by the
sites of the underlying lattice which are occupied by the droplets.
The edges of the graph connect pairs of neighboring occupied vertices on
the lattice. Since each site may be occupied by a droplet, the statistical
ensemble we consider here includes the sub-graphs of the lattice (though
most of them have negligible weight).

We divide each droplet into {\em faces}, one for each edge of the graph
that emerges from its vertex. If another droplet occupies
a vertex that is connected to the same opposite edge, then
the corresponding face is associated with the edge between the two droplets,
and we call it a \emph{linked face}. If there is no neighboring droplet at
the vertex connected to a given edge, we term it a \emph{free face}.

We consider systems in which
the radius of gyration of a polymer is smaller than the radius of
a droplet. Thus, we treat each edge and each free face as a separate system.
In an experiment, where the total number of polymers is fixed, the edges and faces
interact through this constraint. In the multi-canonical ensemble, however,
they do not interact.  We can thus treat each edge or free face as an independent
system and write its partition function explicitly.

As discussed above, excluded volume interactions imply that a  face
can contain a maximum of $\sigma_M$ stickers. This means that an edge has
a maximal occupation of
$\sigma_M$ polymers. These polymers can be links - with the two telechelic stickers
of a given polymer attached
to the two opposing faces of adjacent droplets, or loops - with both stickers
of a given polymers attached to the same face of a single droplet.
A face can contain a maximum of  $\sigma_M/2$ such loops.

\subsection{Polymer free energy}
We consider a model in which,
except for the excluded volume interactions that determine
the maximal number of polymers per droplet, $\sigma_M$,
polymers interact only through their stickers (telechelic ends).
The interactions between the polymer chains have a negligible contribution
for the case of relatively short and stiff chains that we consider here.
Thus the energy difference
between an edge that has two loops on the opposite faces
of two neighboring droplets, and an edge in which these faces are linked by
two polymers, is simply $2\el-2\eb$, where $\el$ is the configurational
free energy of a loop, and $\eb$ is that of a link. We thus define
an effective thermodynamic potential of a polymer in an edge
\begin{equation}\label{eq:mu}
\mu = \mup-\log(e^{-\el}+e^{-\eb}) \;.
\end{equation}
An edge where all the polymers have a thermodynamic potential $\mu$ is equivalent
to a one in which loops have a potential of $\mup+\el$ and links
a thermodynamic potential of $\mup+\eb$,
in the sense that both systems have the same partition function and hence yield
the same distribution $p(\sigma)$. Thus, we assume that all polymers in the edge
have a thermodynamic potential of $\mu$, whether they are loops or links.

In a free face, where the polymers are all in loop configurations, the
thermodynamic potential
of a polymer is $\mup+\el=\mu+\Delta$, where $\Delta=\log(e^{\el-\eb}+1)$.

The free energy of an edge or a free face depends only on the
effective thermodynamic potential
of the polymers and the interaction between the stickers,
which is similar in both the linked faces and free faces. We denote the
free energy of an edge as $\sigma_M\fe(\mu)$, where $\fe(\mu)$ is the free energy
density (free energy per polymer site)
of the face surface, or the free energy of one of the $\sigma_M$
sites on the face.
Since in our model a free face differs from an edge only in the number
of sites available ({\it i.e.}, only $\sigma_M/2$ polymers
can be accommodated on a free face) and in the effective thermodynamic potential,
its free energy is $(\sigma_M/2)\fe(\mu+\Delta)$.

One can readily derive the free energy, $\fe$, of a system with only excluded volume
interactions: $\fe=-\log(1+e^{-\mu})$. We considered such a system in
Subsection \ref{ssev}, where we ignored the configurational free energy of
the polymers.

\begin{figure}
\includegraphics[width=\figurewidth]{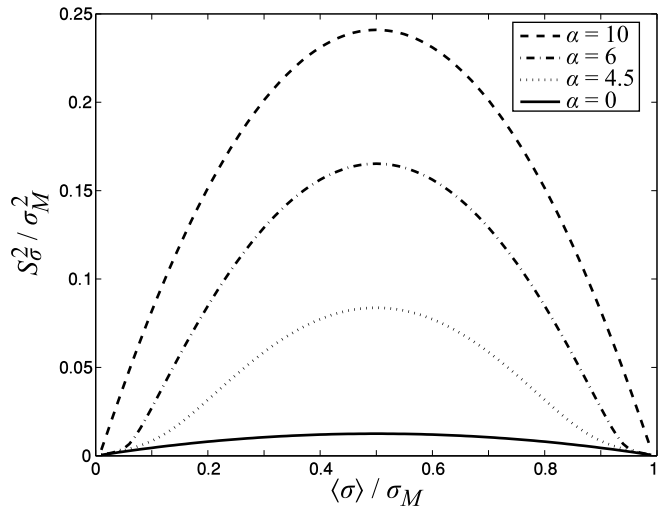}
\caption{The variance of the number of polymers in a bond, as a function
of the average bond occupation (determined by the chemical potential $\mu$),
for $\sigma_M=20$.
The values of the attractive interaction, $\alpha$, are in units of $\kT$.}
\label{figvarsig}
\end{figure}

The interactions between the stickers are accounted for within a mean field
approximation. The total interaction energy for an edge
with $\sigma_M$ polymers is written as $-\alpha\sigma_M/2$.
The number of stickers that occupy a given edge is $\sigma$.
We therefore write the interaction free energy due to two-body interactions
among the stickers
using a virial-type term: $-(\alpha/2\sigma_M)\sigma^2$.
The partition function is
\begin{eqnarray}\label{hs1}
&& Z = \sum_{\sigma=0}^{\sigma_M} \binom{\sigma_M}{\sigma}
\exp\left(-\mu\sigma+\frac{\alpha}{2\sigma_M}\sigma^2\right)
\nonumber\\* &&\ \ \ =
\sqrt{\frac{\sigma_M}{2\pi\alpha}}\int_{-\infty}^{\infty}\upd\phi\left[
e^{-\phi^2/2\alpha} \left(e^{\phi-\mu}+1\right)\right]^{\sigma_M}
\end{eqnarray}
We consider this expression for $Z$ as the partition function of an effective
Hamiltonian,
$H(\phi)=\sigma_M\left[\frac{1}{2\alpha}\phi^2 - \log(e^{\phi-\mu}+1)\right]$,
that has a minimum at $\phi_0$, given by:
\begin{equation}\label{phi0}
\phi_0 = \frac{\alpha}{e^{\mu-\phi_0}+1} \;.
\end{equation}
For $\sigma_M\gg 1$ the distribution of $\phi$ tends to a sharp peak
around $\phi_0$, and we can use the mean field approximation
$\la f(\phi)\ra_\phi\approx f(\phi_0)$ to estimate the average fraction
$\bar\varphi=\la\sigma\ra/\sigma_M$ of sites occupied in an edge and
its variance:
\begin{eqnarray}\label{eq11}
&&\bar\varphi= -\frac{\partial\fe}{\partial\mu}
= \frac{1}{\sigma_M} \left\la \frac{\partial H(\phi)}{\partial\mu} \right\ra_\phi
\approx \frac{\phi_0}{\alpha} \;, \\
&&S^2_{\bar\varphi} = \frac{\partial^2\fe}{\partial\mu^2} \approx
\left( \frac{1}{\bar\varphi(1-\bar\varphi)} - \alpha \right)^{-1} . \label{eqra}
\end{eqnarray}
This expression applies whether $\alpha$ is positive
(attractive interactions), negative (repulsive interactions), or zero.
Eq. \ref{eqra} suggests a divergence of the variance for $\alpha\geq 4$.
In a macroscopic system ($\sigma_M \rightarrow \infty$), this signifies
an instability related to macroscopic phase separation. We do
not expect such behavior for finite $\sigma_M$ (which is more realistic),
but we do observe a pronounced increase in the variance, as the interaction
strength $\alpha$ increases (see Fig. \ref{figvarsig}).

\begin{figure}
\includegraphics[width=\figurewidth]{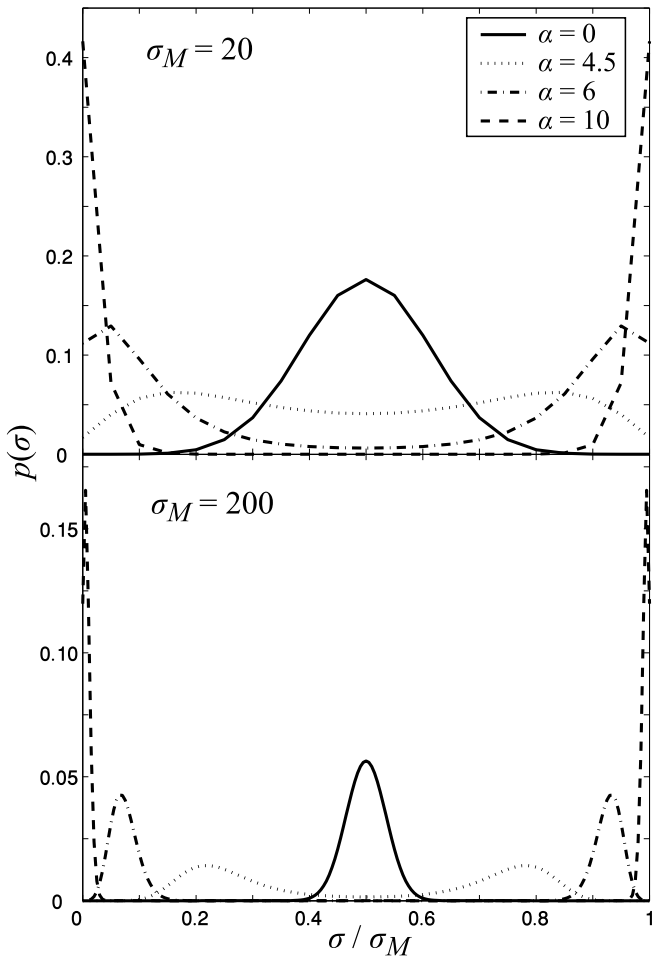}
\caption{The distribution $P(\sigma)$ of the number of polymers in a bond,
for $\sigma_M=20$ and $\sigma_M=200$. The effective chemical potential
$\mu$ is such that $\la\sigma\ra=\sigma_M/2$.
The values of $\alpha$ are in units of $\kT$. }
\label{figbimodal}
\end{figure}

The increase in the variance is due to a bimodal distribution
of the polymer density in the bonds, as demonstrated in Fig. \ref{figbimodal}.
For large values of $\alpha$, the distribution $p(\sigma)$ resembles the bimodal
toy distribution in Eq. \ref{eqtoybin}.

In such a system, when the interactions are strong enough, one may observe two
kinds of edges connecting the droplets: dilute edges containing one or a few
polymers, and dense edges, containing about $\sigma_M$ polymers. Both kinds
induce the same effective attraction between the droplets (see below).
Under degradation, most of the dilute edges will break down, and the connectivity
of the network may decrease dramatically, as depicted in Fig. \ref{fig1}d.
If the system was in the gel phase, and the connectivity is reduced below the
percolation threshold, the gel will transform into a fluid phase.

We conclude by obtaining an analytical expression for the mean-field free
energy of an edge, which is a good approximation for the case $\sigma_M\gg 1$.
Expansion of the effective Hamiltonian around its minimum yields
$H(\phi)\approx H(\phi_0)+ \half H''(\phi_0) (\phi-\phi_0)^2$, where
$H''(\phi) = \partial^2 H / \partial\phi^2$. Plugging this approximation into
the integral of Eq. \ref{hs1} and taking the logarithm, we obtain
the grand-canonical potential for a single edge:
\begin{eqnarray}\label{eq13}
&&\fe(\mu) = -\log(Z) \approx \frac{1}{\sigma_M}\left[ H(\phi_0) -
\half\log\left(\frac{\alpha}{\sigma_M}H''(\phi_0)\right) \right]
\nonumber \\* &&\ =
\frac{\alpha}{2}\bar\varphi^2 + \log(1-\bar\varphi) - \frac{1}{2\sigma_M}
\log(1-\alpha\bar\varphi(1-\bar\varphi)) \;,
\end{eqnarray}
where $\bar\varphi$ is determined by the value of $\mu$, and the last
equality is obtained using Equations \ref{phi0} and \ref{eq11}.

\subsection{Droplet free energy} Each vertex of the
network in our model can occupy one oil droplet. We
define $s_i=0,1$ as the occupation number associated with
vertex $i$ in the lattice. The number of edges in a given
configuration of the system is $N_{\rm E}=\sum_{\la ij\ra}s_i
s_j$, where the sum is over pairs of neighboring sites. The
number of free faces is $N_{\rm F}=\sum_{\la ij\ra}
s_i(1-s_j)+s_j(1-s_i)$. The number of droplets is $N_{\rm
d}=\sum_i s_i$.

Each edge has a free energy, $\sigma_M\fe(\mu)$, and every free face has a free
energy, $(\sigma_M/2)\fe(\mu+\Delta)$. The partition function of the system
containing both polymers (which have
both interactions and configurational entropy)
and droplets (which have translational entropy)
is given by the sum over all droplet configurations:
\begin{eqnarray}\label{Za}
&& Z = \sum_{\{s\}} \exp\left[ N_{\rm d}\mu_{\rm d} + N_{\rm E}\sigma_M\fe(\mu)
+ N_{\rm F}\frac{\sigma_M}{2}\fe(\mu+\Delta) \right] \nonumber\\
&& = \sum_{\{s\}} \exp\left[ N_{\rm d}
\left( \mu_{\rm d} + \frac{\zeta}{2}\sigma_M \fe(\mu+\Delta) \right)
+ N_{\rm E} J(\mu,\Delta) \right] , \nonumber\\
&&\
\end{eqnarray}
where $\mu_{\rm d}$ is the chemical potential of the droplets,
$J(\mu,\Delta) = \fe(\mu)-\fe(\mu+\Delta)$ is the effective interaction
between the droplets, mediated by the polymers, and $\zeta$ is the number of
nearest  neighbors at each site. This is equivalent to the  partition function
of a lattice gas model \cite{Zilman03}.

We use a Landau type, coarse grained approximation for
the free energy, as a function of the dimensionless concentration
(volume fraction) of droplets, $c=\la s\ra$, which is the average occupation
per vertex:
\begin{eqnarray}\label{landau}
&&f = c\log c+(1-c)\log(1-c)
\nonumber\\* && \ \ \ \
 + \half\zeta\sigma_M\fe(\mu+\Delta) c
 + \half\zeta\sigma_M J(\mu,\Delta) c^2 \;.
\end{eqnarray}
The first two terms in Eq. \ref{landau} give the lattice gas entropy;
the third term is obtained using $N_{\rm d}/V=c$, where $V$ is the volume
of the system; the fourth term is obtained from Eq. \ref{Za} using
the mean-field approximation $N_{\rm E}/V \approx \zeta c^2/2$.
The dependence of $f$ on the potential $\mu_{\rm d}$ is linear
in the droplet concentration, and this term has
no effect on the phase behavior of the system.

In the coexistence region, there is an equilibrium of droplet chemical
potential, $\mu_{\rm d}=\partial f/\partial c$, and of the osmotic pressure,
$\pi_{\rm d}=f-\mu_{\rm d} c$, between the coexisting phases.
In equilibrium, the polymer chemical potential must
be the same in both phases; this is automatically ensured by
working at constant $\mu$. By solving these equalities we obtain
the binodal curve in $c\times\mu$ (droplet concentration-polymer
thermodynamic potential) plane.

Since in experiment it is easier to control the polymer
concentration $\varphi$ than the polymer chemical
potential, it is useful to predict the phase diagram in the
$c\times\varphi$ plane. We perform this transformation using the
identity
\begin{equation}\label{eqvarphi}
\varphi = \frac{\upd f}{\upd\mu} = \sigma_M \left[
\fe'(\mu+\Delta) c + J'(\mu,\Delta) c^2 \right] \;,
\end{equation} where the prime indicates a
derivative with respect to $\mu$.
This fixes the polymer chemical potential, $\mu$, as a function
of the number of polymers, $\varphi$. In the
phase-coexistence region, above the binodal curve, the
conditions of constant droplet chemical potential, $\mu_{\rm
d}$, and osmotic pressure, $\pi_{\rm d}$, for fixed polymer
chemical potential, $\mu$, yield values of the droplet density,
$c$, for the coexisting dilute and dense phases respecitvely.
From Eq. \ref{eqvarphi} this implies that
each phase has a different value of the polymer
concentration, $\varphi$.

The critical point in the $c\times\mu$ (droplet concentration-polymer
concentration) plane occurs at $c=1/2$, with $\mu$
given by
\begin{equation}\label{critical}
J(\mu,\Delta) = \frac{-4}{\zeta\sigma_M} \;.
\end{equation}

\begin{figure} \includegraphics[width=\figurewidth]{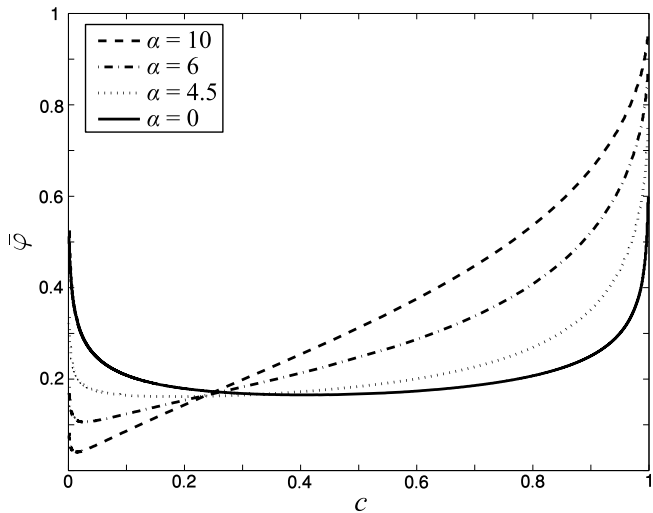}
\caption{The phase coexistence curves of systems
with different values for the interaction between the stickers, $\alpha$
(given in units of $\kT$).
The curves are obtained from the condition of
equilibrium of the droplet chemical potentials, $\mu_{\rm d}$,
and  osmotic pressure, $\pi_{\rm d}$, in the coexisting phases,
using the Landau free energy is given by Eq. \ref{landau}. The
$x$-axis represents the droplet volume fraction, $c$, and the
$y$-axis represents the number of polymers per available site,
$\bar\varphi$. The parameters used are $\Delta=0.3\kT$,
$\zeta=4$ and $\sigma_M=20$.} \label{figpd} \end{figure}

The binodal curve of the system is presented in Fig. \ref{figpd}
for different values of the edge interaction $\alpha$. The
maximal number of stickers that can reside in one
droplet is $\zeta\sigma_M$. Thus, the number of stickers per
unit volume is  $c\zeta\sigma_M$, and the maximal number of
polymers per unit volume is $c\zeta\sigma_M/2$ (assuming a
prohibitively high energetic cost of a hydrophobic sticker in
the water). The number of polymers per available site in the
system is then
\begin{equation} \label{eqbarvp}
\bar\varphi = \frac{2 \varphi}{\zeta \sigma_M c} \;.
\end{equation}

A phase co-existence region is present in a system
without explicit sticker interactions ($\alpha=0$)
\cite{Zilman03}. However, the interactions can markedly affect
the shape of the coexistence curve and of the concentrations in
the two co-existing phases. As the interaction, $\alpha$,
increases, the effective interaction between the droplets,
induced by the edges, increases and becomes more sensitive to
the polymer concentration. As a result, the onset of phase
separation is sharper, and characterized by a phase with  very
dilute droplets and low polymer concentration that coexists with
a phase with high droplet density and high polymer concentration.
In fact, in Fig. \ref{figpd} the  curve for $\alpha=10$ is almost
parallel to the tie-lines.

We note that dilute edges (with one or a few polymers) can also exist
in the dense droplet phase, since they induce the same effective
interaction, $J(\mu,\Delta)$, between the droplets as the dense edges, and their
polymers have the same thermodynamic potential, $\mu$, as the polymers
in the dense edges. This scenario may change if
we consider interactions between the edges.

\section{Discussion}

In this paper, we have predicted the immunity of the gel phase of
a polymer-microemulsion system to random degradation of polymers.
The gel in this system consists of oil droplets connected by
telechelic polymers.   The set of polymers that link
a given pair of droplets is defined as an edge ({\it{i.e.}},
bond) of the gel network. We consider the distribution,
$p(\sigma)$, of the number of polymers in one such edge, and
argue that immunity of the gel is inversely related to the
variance, $S^2_\sigma$, of this distribution.

Repulsive interactions between the stickers (the telechelic ends of
the polymers) reduce the variance $S^2_\sigma$, thus contributing to the gel
immunity. Attractive interaction, which increase $S^2_\sigma$,
reduce the immunity of the gel.

Attractive interactions may give rise to a bimodal distribution of the
edge occupation, which can give rise to the co-existence of dense and dilute
edges in the same macroscopic sample. A dilute edge contains a small number
of polymers, and is likely to vanish under a random degradation of the polymers.
When a finite fraction (say, $1/2$) of the edges are dilute, the fraction
of edges surviving a degradation will be lower, and the system is more likely
to transform from the gel to the fluid phase.

This reasoning is particularly applicable to the case
of a dense phase that coexists with a dilute phase (assuming
the dilute phase is not a gel). Since the polymer density
\emph{within} the dense droplet phase is high, we expect -- if the
attractive interactions are large enough --
to find a bimodal distribution of the edge occupation within the
dense droplet phase. In that case, we are interested in the
immunity of the dense phase, and therefore must take into
account the distribution, $p(\sigma)$, within the dense droplet
phase.

To test our predictions for the immunity of a system with
attractive interactions among the polymers, we suggest
comparing   the elastic stability (gel-like nature) with
respect to polymer degradation
of  two systems (both in the dense phase):  one with no sticker
interactions ($\alpha=0$) and one with strong interactions (say,
$\alpha=10$). According to our
predictions,  the system with no interactions will  be
characterized by  edges each of which has approximately the same
number of polymers,  while the system with strong interactions
will show a bimodal distribution for the edge occupancy. We
predict that after  polymer degradation the fraction of
surviving edges will be larger in the non-interacting system
(see Fig. \ref{fig1}), and that this system is more likely to
remain in the gel phase.

In order to investigate the immunity of such gels experimentally,
it is necessary to devise a method to degrade the polymers
at random and in a homogeneous manner throughout the system.
It may be difficult to do this using molecules in a  solvent,
since a solvent added to the system will initially have
a much higher concentration at the surface of the gel (penetrating the bulk by
slow diffusion); thus the polymers
near the surface are much more likely to be degraded.

One way to achieve random degradation may be to use
polymers that have been synthesized with
a photo-sensitive molecule in the middle of each chain.
Optical methods might be used to cause these molecules
to change their conformation and depolymerize.
The intensity and duration of the optical beam can be used to control
the fraction of polymers broken.

\begin{acknowledgments}
The authors thank L. Chai, M. Gottlieb, S. Havelin,
  J. Klein, C. Ligoure, G. Porte,  and A. Zilman for useful
discussions. This work has been supported by the German Israeli
Foundation and by an EU Network Grant.
\end{acknowledgments}


\begin{thebibliography}{6}
\bibitem{AlbertRMP}
R. Albert and A. L. Barab\'asi, Rev. Mod. Phys. \textbf{74}, 47 (2002).
\bibitem{Wagner00}
A. Wagner, Nature Genetics \textbf{24}, 355 (2000).
\bibitem{Albert00}
R. Albert, H. Jeong and A. L. Barab\'asi, Nature \textbf{406}, 378 (2000).
\bibitem{Cohen03}
R. Cohen R, S. Havlin and D. Ben-Avraham, Phys. Rev. Lett. \textbf{91},
247901 (2003).
\bibitem{Porte}
E. Michel, J. Appell, F. Molino, J. Kieffer and G. Porte,
J. Rheol. \textbf{45}, 1465 (2001);
M. Filali, M. J. Ouazzani, E. Michel, R. Aznar, G. Porte and J. Appel,
J. Phys. Chem. B \textbf{105}, 10528 (2001).
\bibitem{Gottlieb}
M. Adam, D. Lairez, M. Karpasas and M. Gottlieb, Macromolecules \textbf{30},
5920 (1997).
\bibitem{Zilman03}
A. Zilman, J. Kieffer, F. Molino, G. Porte and S. A. Safran,
Phys. Rev. Lett. \textbf{91}, 015901 (2003).
\end{thebibliography}
\end{document}